\documentclass{llncs}

\begin{document}
\pagestyle{empty}

\mainmatter

\title{Time and Space Bounds for 
Reversible Simulation\thanks{All authors are partially supported by the
EU fifth framework project QAIP, IST--1999--11234,
the NoE QUIPROCONE IST--1999--29064,
the ESF QiT Programmme, and the EU Fourth Framework BRA
 NeuroCOLT II Working Group
EP 27150.
Buhrman and Vit\'anyi are also affiliated with
the University of Amsterdam. Tromp is also affiliated with
Bioinformatics Solutions, Waterloo, N2L 3G1 Ontario, Canada.
}
\\
{\small ({\it Extended Abstract}) }}
\titlerunning{Lecture Notes in Computer Science}
\author{Harry Buhrman \and John Tromp \and
Paul Vit\'{a}nyi}

\authorrunning{Buhrman, Tromp, and Vit\'{a}nyi}

\institute{
CWI, Kruislaan 413, 1098 SJ Amsterdam,\\
The Netherlands, email{\{buhrman,tromp,paulv\}@cwi.nl} 
}
 
\maketitle
 
\begin{abstract}
We prove a general upper bound on the tradeoff between time and
space that suffices for the reversible simulation of irreversible
computation. Previously, only simulations
using exponential time or quad\-ra\-tic space were known.
 The tradeoff shows for the first time that we can simultaneously
achieve subexponential time and subquadratic space.
 The boundary values are the exponential time 
with hardly any extra space required by the Lange-McKenzie-Tapp method
and the ($\log 3$)th power time with square space required by
the Bennett method.
We also give the first general
lower bound on the extra storage space required
by general reversible simulation. This lower bound is optimal
in that it is achieved by some reversible simulations.
\end{abstract}

\section{Introduction}
Computer power
has roughly doubled every 18 months for the last half-century
(Moore's law).
This increase in power is due primarily to the continuing
miniaturization of the elements of which computers are made,
resulting in more and more elementary gates per unit area
 with higher and higher
clock frequency, accompanied by less and less
energy dissipation per elementary computing event. Roughly, a
linear increase in clock speed is accompanied by a square increase
in elements per unit area---so if all elements
compute all of the time, then the dissipated energy per time unit rises
cubicly (linear times square)
in absence of energy decrease per elementary event. The continuing
dramatic decrease in dissipated energy per elementary event is
what has made Moore's law possible. But there is a foreseeable
end to this: There is a
minimum quantum of energy dissipation associated with
elementary events. This puts a fundamental limit
on how far we can go with miniaturazation, or does it?

{\bf Reversible Computation:}
R. Landauer~\cite{La61}
has demonstrated
that it is only the `logically
irreversible' operations in a physical computer
that necessarily dissipate energy by generating a
corresponding amount of entropy for every bit of information
that gets irreversibly erased; the logically reversible operations
can in principle be performed dissipation-free.
Currently, computations are commonly irreversible, even though the
physical devices that execute them are fundamentally reversible.
At the basic level, however, matter is governed by classical
mechanics and quantum mechanics,
which are reversible.
This contrast is only possible at the cost of efficiency loss
by generating thermal entropy into the environment.
With computational device technology rapidly approaching
the elementary particle level it has been
argued many times that this effect gains in significance
to the extent that
efficient operation (or operation at all)
of future computers requires them to
be reversible (for example, in
\cite{La61,Be73,Be82,FT82,Ke88,LV96,FKM97}).
The mismatch of computing organization and reality
will express itself in friction: computers will 
dissipate a lot of heat unless their mode of operation becomes
reversible, possibly quantum mechanical. 
Since 1940 the
dissipated energy per bit operation in a computing
device has---with remarkable regularity---decreased
at the inverse rate of Moore's law \cite{Ke88} (making Moore's law possible).
Extrapolation of current trends
shows that the energy dissipation per
binary logic operation needs to be reduced below $kT$
(thermal noise)
within 20 years. Here $k$ is Boltzmann's constant and $T$
the absolute temperature in degrees Kelvin,
so that $kT \approx 3 \times 10^{-21}$
Joule at room temperature. Even at $kT$ level,
a future device containing 1 trillion ($10^{12}$) gates
operating at 1 terahertz ($10^{12}$) switching all gates all of the
time dissipates about 3000 watts.
Consequently, in contemporary
 computer and chip architecture design the issue of power consumption
has moved from a background worry to a major problem.
For current research towards implementation of reversible computing
on silicon see MIT's Pendulum Project and linked web pages
(http://www.ai.mit.edu/$\sim$cvieri/reversible.html).
On a more futuristic note,
quantum computing \cite{Sh94,NC00} is reversible.
Despite its importance, theoretical advances in reversible
computing are scarce and far between; all serious ones are listed 
in the references.

{\bf Related Work:}
Currently, almost no algorithms and
other programs are designed according to reversible principles
(and in fact, most tasks like computing Boolean functions are inherently
irreversible). To write reversible programs by hand is
unnatural and difficult. The natural way is to
compile irreversible
programs to reversible ones. This raises the question about
efficiency of general reversible simulation of irreversible
computation. Suppose the irreversible computation to be
simulated uses $T$ time and $S$ space.
 A first efficient method was proposed by Bennett \cite{Be89},
but it is space hungry and uses
\footnote{By
judicious choosing of simulation parameters this method can be tweaked to
run in $ST^{1+\epsilon}$ time for every $\epsilon >0$
at the cost of introducing a multiplicative constant depending
on $1/\epsilon$. The complexity analysis of \cite{Be89} 
was completed in \cite{LeSh90}.
} 
 time $ST^{\log 3}$ and
space
$S \log T$. If $T$ is maximal, that is, exponential in $S$,
then the space use is $S^2$. This method can be modelled by a reversible
pebble game. Reference \cite{LTV98} demonstrated
that Bennett's method is optimal
for reversible pebble games and that simulation space can be traded
off against limited erasing.
In \cite{LMT97} it was shown that using a method by Sipser \cite{Si}
one can reversibly simulate using only $O(S)$ extra space but at the cost
of using exponential time.
In \cite{FA98} the authors provide an oracle construction (essentially
based on
\cite{LTV98}) that separates reversible and irreversible
space-time complexity classes.

{\bf Results:}
Previous results seem to suggest that a reversible simulation is stuck
with either quadratic space use or exponential time use. 
This impression turns out to be false: 
\footnote{
The work reported in this paper
dates from 1998; 
Dieter van Melkebeek has drawn our attention
to the unpublished \cite{W00} 
with similar, independent but later, research.
}

Here we prove a
tradeoff between time and space which has the exponential time
simulation and the quadratic space simulation as extremes and
for the first time gives a range of simulations using simultaneously
subexponential 
($2^{f(n)}$ is subexponential if $f(n) = o(n)$) time and subquadratic space.
The idea is to use Bennett's pebbling game where the pebble steps
are intervals of the simulated computation that are bridged by
using the exponential simulation method. (It should be noted
that embedding Bennett's pebbling game in the exponential method
gives no gain, and neither does any other iteration of embeddings
of simulation methods.) Careful analysis
shows that the simulation using $k$ pebbles takes
$T' := S 3^k 2^{O(T/2^{k})}$ time and $ S' = O(kS)$ space, and
in some cases the upper bounds are tight. For $k=0$ we
have the exponential time simulation method and for $k= \log T$
we have Bennett's method. Interesting values arise for say

(a) $k= \log \log T$: $T'= S (\log T)^{\log 3} 2^{O(T/\log T)}$
and $S'= S \log \log T  \leq S \log S$; 

(b) $k= \sqrt{\log T}$:
$S' = S \sqrt{\log T} \leq S \sqrt{S}$ and 
$T' = S 3^{\sqrt{\log T}} 2^{O(T/2^{\sqrt{\log T}})}$.

(c)  
Let $T,S,T',S'$ be as above. 
Eliminating the unknown $k$ shows the tradeoff
between simulation time $T'$ and extra
simulation space $S'$:
$ T' = S3^{\frac{S'}{S}}2^{O(T/2^{{\frac{S'}{S}}})} $.

(d) Let $T,S,T',S'$ be as above and let the irreversible
computation be halting and compute a function
from inputs of $n$ bits to outputs. For general reversible simulation 
by a reversible Turing machine
using a binary tape alphabet and a single tape,
$S' \geq n + \log T + O(1)$ and $T' \geq T$.
This lower bound is optimal in the sense that
it can be achieved by simulations
at the cost of using time exponential in $S$.

{\bf Main open problem:}
The ultimate question is whether one can
do better, and obtain improved upper and lower bounds on
the tradeoff between time and space of reversible simulation,
 and in particular whether one can have almost linear time
and almost linear space simultaneously.  

\section{Reversible Turing Machines}
In the standard model of a Turing machine
the elementary operations are rules
in quadruple format $(p,s,a,q)$ meaning that
if the finite control is in state $p$ and the machine
scans tape symbol $s$, then the machine performs action $a$
and subsequently the finite control enters state $q$.
Such an action $a$ consists of either printing a symbol $s'$
in the tape square scanned, or moving the scanning head
one tape square left or
right.

Quadruples are said
to {\em overlap in domain} if they cause the machine in the same state
and scanning the same symbol to perform different actions.
A {\em deterministic Turing machine} is defined as a Turing machine
with quadruples no two of which overlap in domain.

Now consider the special format (deterministic) Turing
machines using quadruples of two
types: {\em read/write} quadruples and {\em move} quadruples.
A read/write quadruple $(p,a,b,q)$ causes the machine in state
$p$ scanning tape symbol $a$ to write symbol $b$ and enter state $q$.
A move quadruple $(p,\ast,\sigma ,q)$ causes the machine
in state $p$ to move its tape head by $\sigma \in \{-1,+1\}$
squares and enter state $q$,
oblivious to the particular symbol in the currently scanned tape square.
(Here `$-1$' means `one square left', 
and `$+1$' means `one square right'.) Quadruples are said
to {\em overlap in range} if they cause the machine to enter
the same state and either both write the same symbol or
(at least) one of them moves the head. Said differently,
quadruples that enter the same state overlap in range
unless they write different symbols.
A {\em reversible Turing machine} is a deterministic Turing machine
with quadruples no two of which overlap in range.
A $k$-tape reversible Turing machine uses $(2k+2)$ tuples
which, for every tape separately, select a read/write or move on that
tape. Moreover, any two tuples can be restricted to some single tape
where they don't overlap in range.

To show that every partial recursive function can be computed
by a reversible Turing machine one can proceed as follows \cite{Be73}.
Take the standard irreversible Turing machine computing that function.
We modify it by
adding an auxiliary storage tape called the `history tape'.
The quadruple rules are extended to 6-tuples to additionally
manipulate the history tape.
To be able to reversibly undo (retrace)
the computation deterministically, the new 6-tuple
rules have the effect that the machine keeps a record
on the auxiliary history tape consisting of
the sequence of quadruples executed on the original tape.
Reversibly undoing a computation
entails also erasing the record
of its execution from the history tape.
This notion of reversible computation means
that only $1:1$ recursive functions can be computed.
To reversibly simulate an
irreversible computation from $x$ to $f(x)$
one reversibly  computes from input $x$
to output $\langle x, f(x) \rangle$.

Reversible Turing machines or other reversible computers
will require special reversible programs. One feature of such
programs is that they should be executable when read from bottom
to top as well as when read from top to bottom. Examples are
the programs $F(\cdot)$ and $A(\cdot)$ in \cite{LTV98}.
In general, writing reversible programs will be difficult.
However, given a general reversible simulation of irreversible computation,
one can simply write an oldfashioned
irreversible program in an irreversible programming language,
and subsequently simulate it reversibly. This leads to the following:
\begin{definition}
An {\em irreversible-to-reversible compiler}
receives an irreversible program as input and
 compiles
it to a reversible program. 
\end{definition}
Note that there is a decisive difference between reversible circuits and
reversible special purpose computers \cite{FT82} on the one hand, and
reversible universal computers on the other hand \cite{Be73,Be89}.
While one can design a special-purpose reversible version for every particular
irreversible circuit using reversible universal gates, such a method
does not yield an irreversible-to-reversible compiler
that can execute any irreversible
program on a fixed universal reversible computer architecture as
we are interested in here.

\section{Time Parsimonious Simulation}

\subsection{Background}
We keep the discussion at an intuitive informal level; the
cited references contain the formal details and rigorous constructions.
An irreversible deterministic Turing machine has an infinite graph of {\em all}
configurations where every configuration has outdegree at most one. In a 
reversible deterministic Turing machine every configuration also
has indegree at most one. The problem of reversing an irreversible
computation from its output is to revisit the input configurations
starting from the output configuration by a process of reversibly
traversing the graph.

The reversible Bennett strategy \cite{Be89} 
essentially reversibly visits only the
linear graph of configurations visited by the irreversible deterministic
Turing machine in its computation from input to output, and no other
configurations in the graph. It does so by a recursive procedure 
of establishing and undoing intermediate checkpoints that are kept
simultanously in memory. It turns out that this can be done 
using limited time $T^{\log 3}$ and space $S \log T$.

\subsection{Reversible Pebbling}
Let $G$ be a linear list of
nodes $\{1,2, \ldots , T_G \}$.
We define a {\em pebble game} on $G$ as follows. The game
proceeds in a discrete sequence of steps of a single {\em player}.
There
are $n$ pebbles which can be put on nodes of $G$.
At any time the set of pebbles is divided in
pebbles on nodes of $G$ and the remaining pebbles which are called
{\em free} pebbles. At every step either an existing
 free pebble can be put
on a node of $G$ (and is thus removed from the free pebble pool)
 or be removed from a node of $G$ (and is added to the
free pebble pool).
Initially $G$ is unpebbled and there is a pool of free pebbles.
The game is played according to the following rule:

\begin{description}
\item[Reversible Pebble Rule:]
If node $i$ is occupied by a pebble, then one may either
place a free pebble on node $i+1$ (if it was not occupied before), or
remove the pebble from node $i+1$.
\end{description}

We assume an extra initial node $0$ permanently
occupied by an extra, fixed pebble,
so that node $1$ may be (un)pebbled at will.
This pebble game is inspired by the method of simulating irreversible Turing
Machines on reversible ones in a space efficient manner. The placement
of a pebble corresponds to checkpointing the next state of the irreversible
computation, while the removal of a pebble corresponds to reversibly erasing
a checkpoint. Our main interest is in determining the number of pebbles $k$
needed to pebble a given node $i$.

The maximum number $n$ of pebbles
which are simultaneously on $G$
at any one time in the game gives the space complexity
$nS$ of the simulation. If one deletes a pebble not following
the above rules, then this means a block of bits of size $S$ is
erased irreversibly. 

\subsection{Algorithm}

\label{reachable}

We describe the idea of Bennett's simulation \cite{Be89}.
This simulation is optimal \cite{LTV98} among all reversible pebble games.
The total computation of $T$ steps is broken into $2^k$ segments of
length $m=T2^{-k}$.
Every $m$th point of the computation is a node in the pebbling game;
node $i$ corresponding to $im$ steps of computation.

For each pebble a section of tape is reserved long enough to store the
whole configuration of the simulated machine. By enlarging the tape
alphabet, each pebble will require space only $S+O(1)$.


Both the pebbling and unpebbling of a pebble $t$ on some node,
given that the previous node has a pebble $s$ on it, will be achieved
by a single reversible procedure bridge($s,t$). This looks
up the configuration at section $s$, simulates $m$ steps of computation
in a manner described in section~\ref{bridge}, and exclusive-or's the result
into section $t$.
If $t$ was a free pebble, meaning that its tape section is all zeroes,
the result is that pebble $t$ occupies the next node. If $t$ already
pebbled that node then it will be zeroed as a result.

The essence of Bennett's simulation is a recursive subdivision
of a computation path into 2 halves, which are traversed in 3 stages;
the first stage gets the midpoint pebbled, the second gets the
endpoint pebbled, and the 3rd recovers the midpoint pebble.
The following recursive procedure implements this scheme;
Pebble($s,t,n$) uses free pebbles $0,\ldots,n-1$ to compute
the $2^n$th node after the one pebbled by $s$,
and exclusive-or's that node with pebble $t$
(either putting $t$ on the node or taking it off).
Its correctness follows by straightforward induction.
Note that it is its own reverse;
executing it twice will produce no net change.
The pebble parameters $s$ and $t$ are simply numbers
in the range $-1,0,1,\ldots,k$. Pebble -1 is permanently on node 0,
pebble $k$ gets to pebble the final node, and pebble $i$,
for $0\leq i < k$ pebbles nodes that are odd multiples of $2^i$.
The entire simulation is carried out with a call pebble($-1,k,k$).


\begin{tabbing}
pe\=bb\=le($s,t,n$) \\
\{ \\
\> if ($n=0$) \\
\> \> bridge($s,t$); \\
\> fi ($n=0$) \\
\> if ($n>0$) \\
\> let $r=n-1$ \\
\> pebble($s,r,n-1$); \\
\> pebble($r,t,n-1$); \\
\> pebble($s,r,n-1$) \\
\> fi ($n>0$) \\
\}
\end{tabbing}

As noted by Bennett, both branches and merges must be labeled with
mutually exclusive conditions to ensure reversibility.
Recursion can be easily implemented reversibly by introducing an extra
stack tape, which will hold at most $n$ stack frames of size $O(\log n)$ each,
for a total of $O(n \log n)$.

This pebbling method is optimal in that no more
than $2^{n+1}-1$ steps can be bridged with $n$ pebbles \cite{LTV98}.
A call pebble($s,t,n$) results in $3^n$ calls to bridge($\cdot$,$\cdot$).
Bennett chose the number of pebbles large enough ($n=\Omega(\log T)$)
so that $m$ becomes small, on the order of the space $S$ used by the simulated
machine. In that case bridge($s,t$) is easily implemented with the help
of an additional {\em history} tape of size $m$ which records the
sequence of transitions. Instead, we allow an arbitrary choice of $n$
and resort to the space efficient simulation of \cite{LMT97} to
bridge the pebbled checkpoints.

\section{Space Parsimonious Simulation}
\label{bridge}
Lange, McKenzie and Tapp, \cite{LMT97}, devised a reversible simulation,
{\em LMT-simulation} for short,
that doesn't use extra space, at the cost of using exponential time.
Their main idea of reversibly simulating a machine without using more space
is by reversibly cycling through the configuration tree of the machine
(more precisely the connected component containing the input configuration).
This configuration tree is a tree whose nodes
are the machine configurations and where two nodes are connected
by an edge if the machine moves in one step from one configuration
to the other. We consider each edge to consist of two {\em half-edges},
each adjacent to one configuration.

The configuration tree can be traversed by alternating
two permutations on half-edges: a swapping permutation which
swaps the two half-edges constituting each edge, and a rotation
permutation whose orbits are all the half-edges adjacent to one
configuration. Both permutations can be implemented in a constant number
of steps. For simplicity one assumes the simulated machine
strictly alternates moving and writing transitions.
To prevent the simulation from exceeding the available space $S$,
each pebble section is marked with special left and right
markers $\dagger, \ddagger$, which we assume the simulated machine not
to cross. Since this only prevents crossings in the forward simulation,
we furthermore, with the head on the left (right) marker,
only consider previous moving transitions from the right (left).

\section{The Tradeoff Simulation}
To adapt the LMT simulation to our needs, we equip our simulating machine
with one extra tape to hold the simulated configuration and
another extra tape counting the difference between forward and backward
steps simulated. $m=2^n$ steps of computation can be bridged with a $\log m$
bits binary counter, incremented with each simulated forward step,
and decremented with each simulated backward step---
incurring an extra $O(\log m)$ factor slowdown in simulation speed.
Having obtained the configuration $m$ steps beyond that of pebble $s$,
it is exclusive-or'd into section $t$ and then the LMT simulation is
reversed to end up with a zero counter and a copy of section $s$, which
is blanked by an exclusive-or from the original.

\begin{tabbing}
br\=id\=ge\=($s,t$) \\
\{ \\
\> copy section $s$ onto (blanked) simulation tape \\
\> setup: goto enter; \\
\> loop1: come from endloop1; \\
\> simulate step with swap\&rotate and adjust counter \\
\> if (counter=0) \\
\> \> rotate back; \\
\> \> if (simulation tape = section $s$) \\
\> \> \> enter: come from start; \\
\> \> fi (simulation tape = section $s$) \\
\> fi (counter=0) \\
\> endloop1: if (counter!=$m$) goto loop1; \\
\> exclusive-or simulation tape into section $t$ \\
\> if (counter!=$m$) \\
\> \> loop2: come from endloop2; \\
\> reverse-simulate step with anti-rotate\&swap and adjust counter \\
\> if (counter=0) \\
\> \> rotate back; \\
\> \> if (simulation tape = section $s$) goto exit; \\
\> fi (counter=0) \\
\> endloop2: goto loop2; \\
\> exit: clear simulation tape using section $s$ \\
\}
\end{tabbing}

\subsection{Complexity Analysis}
Let us analyze the time and space used by this simulation.
\begin{theorem}
An irreversible computation using time $T$ 
and space $S$ can be simulated reversibly
in time $T'=  3^k 2^{O(T/2^k)}S$ and space $S' = S(1+O(k))$,
where $k$ is a parameter that can be chosen freely $0 \leq k \leq \log T$
to obtain the required tradeoff between reversible time $T'$
and space $S'$.
\end{theorem}
\begin{proof} 
(Sketch)
Every invocation of the bridge() procedure
 takes time $O(2^{O(m)}S)$.
That is, every configuration has at most $O(1)$ predecessor
configurations where it can have come from (constant number of states,
constant alphabet size and choice of direction).
Hence there are
$\leq 2^{O(m)}$ configurations to be searched
and about as many potential start configurations leading in $m$ moves
to the goal configuration, and every tape section comparison takes
time $O(S)$.
The pebbling game over $2^k$ nodes takes $3^k$ (un)pebbling steps
each of which is an invocation of bridge().
Filling in $m=T/2^k$ gives the claimed time bound.
Each of the $k+O(1)$ pebbles takes space $O(S)$,
as does the simulation tape and the counter,
giving the claimed total space.
$\qed$
\end{proof}

It is easy to verify that
for some simulations the upper bound
is tight.
The boundary cases, $k=0$ gives the LMT-simulation using exponential
time and no extra space, and $k= \log T$ gives Bennett's simulation
using at most square space and subquadratic time. Taking intermediate
values of $k$ we can choose to reduce time at the cost of
an increase of space use and vice versa.
In particular, special values $k= \log \log T$ and
$k= \sqrt{T}$ give the results using simultaneously subexponential
time and subquadratic space exhibited in the introduction.
Eliminating $k$ we obtain:

\begin{corollary}
Let $T,S,T',S'$ be as above. Then there is a reversible simulation that
has the following tradeoff between simulation time $T'$ and extra
simulation space $S'$:
\[ T' = S3^{\frac{S'}{S}}2^{O(T/2^{{\frac{S'}{S}}})} . \]
\end{corollary}

\subsection{Local Irreversible Actions}
Suppose we have an otherwise reversible computation containing
local irreversible actions. Then we need to reversibly simulate
only the subsequence of irreversible steps, leaving the connecting reversible
computation segments unchanged.  That is, an irreversiblity 
parsimonious
computation is much cheaper to reversibly simulate than an
irreversibility hungry one.

\subsection{Reversible Simulation of Unknown Computing Time}
In the previous analysis we have tacitly assumed that the
reversible simulator knows in advance the number of steps $T$ taken
by the irreversible computation to be simulated.
In this context one can distinguish on-line computations
and off-line computations to be simulated. On-line computations
are computations which interact with the outside environment
and in principle keep running forever. An example
is the operating system of a computer.
Off-line computations are computations which compute a definite
function from an input (argument) to an output (value). For example,
given as input a positive integer number, compute as output all
its prime factors. For every input such an algorithm will have a definite
running time.

There is a well-known simple device 
to remove this dependency for batch computations
 without increasing the simulation time  (and space)
too much. Suppose we want to simulate a computation with
unknown computation time $T$. Then we simulate $t$ steps
of the computation with $t$ running through the sequence
of values $2,2^{2},2^{3}, \ldots$ For every value $t$ takes on
we reversibly simulate the first $t$ steps of
the irreversible computation. If $T>t$ then the computation is not finished
at the end of this simulation. Subsequently we reversibly undo
the computation until the initial state is reached again, set $t:=2t$
and reversibly simulate again. This way we continue until $t\geq T$
at which bound the computation finishes. The total time spent
in this simulation is 
\[ T''  \leq  2 \sum_{i=1}^{\lceil \log T \rceil} 
 S3^{\frac{S'}{S}}2^{O(2^{i - \frac{S'}{S}})}
 \leq 2 T' .\]

\section{Lower Bound on Reversible Simulation}\label{sect.lb}
It is not difficult to show a simple lower bound on the extra
storage space required for general
reversible simulation. We consider only irreversible computations
that are halting computations performing a mapping
from an input to an output. For convenience 
we assume
that the Turing machine has a single
binary work tape 
delemited by
markers $\dagger, \ddagger$ that are placed $S$ positions apart.
Initially the binary input of length $n$
is written left adjusted on the work tape.
At the end of the computation the output
is written left adjusted on the work tape.
The markers are never moved.
Such a machine clearly can perform every computation as long as $S$
is large enough with respect to $n$.
Assume that the reversible simulator is a similar model albeit reversible.
The average number of steps
in the computation is the uniform average over all equally likely
inputs of $n$ bits.
\begin{theorem}\label{theo.lb}
To generally simulate an irreversible halting computation of
a Turing machine as above using storage
space $S$ and $T$ steps on
average, on inputs of length $n$,
by a general reversible computation
using $S'$ storage space 
and $T'$ steps on average, 
the reversible simulator Turing machine
having $q'$ states, requires trivially $T' \geq T$ and
$S'   \geq n + \log T - O(1)$ up to a logarithmic additive term.
\end{theorem}

\begin{proof}
There are $2^n$ possible inputs to the irreversible computation,
the computation on every input using on average $T$ steps.
A general simulation of this machine cannot use the semantics
of the function being simulated but must simulate every
step of the simulated machine. Hence $T' \geq T$.
 The simulator being reversible
requires different configurations for every step of everyone
of the simulated computations
that is, at least $2^nT$ configurations. 
The simulating machine has not more than $ q'2^{S'} S'$ distinct
configurations---$2^{S'}$ distinct values
on the work tape, $q'$ states, 
and $S'$ head positions for the combination
of input tape and work tape. Therefore, 
$ q'2^{S'} S' \geq 2^nT$. That is,
$q'S'2^{S'-n} \geq T$ which shows that $S' - n - \log S' \geq \log T - \log q'$.
$\qed$
\end{proof}

For example, consider irreversible computations that don't use
extra space apart from the space to hold the input, that is, $S=n$. An
example is the computation of $f(x)=0$. 
\begin{itemize}
\item
If $T$ is polynomial in $n$ then $S' = n+ \Omega (\log n)$.
\item
If $T$ is exponential in $n$ then $S' = n+ \Omega (n)$.
\end{itemize}

Thus, in some cases the LMT-algorithm is required
to use extra space if we
deal with halting computations computing a function from
input to output. In the final version of the paper \cite{LMT97} the authors
have added that their simulation uses some extra space for counting
(essentially $O(S)$) in case we require halting computations
from input to output, matching the lower bound above for $S=n$ since their 
simulation uses on average $T'$ steps exponential in $S$. 

{\bf Optimality and Tradeoffs:} 
The lower bound of Theorem~\ref{theo.lb} is optimal in the following sense.
As one extreme, the LMT-algorithm of \cite{LMT97}
discussed above uses $S' = n+ \log T$  space
for simulating irreversible computations of total functions on
inputs of $n$ bits, but
at the cost of using $T' = \Omega ( 2^S)$ simulation time. As the other
extreme, Bennett's simple algorithm in \cite{Be73} uses $T' = O(T)$
reversible simulation time, but at the cost of using $S' = \Omega (T)$
additional storage space. This implies that improvements in determining
the complexity of reversible simulation must consider time-space tradeoffs.

\end{document}